\documentstyle[11pt,aaspp]{article}
\tighten
\singlespace

\def\Min{${}^{\prime}$\llap{.}}
\def\ltsima{$\; \buildrel < \over \sim \;$}
\def\simlt{\lower.5ex\hbox{\ltsima}}
\def\gtsima{$\; \buildrel > \over \sim \;$}
\def\simgt{\lower.5ex\hbox{\gtsima}}
\def\etal{{\it et al.\ }}
\def\cf{{\it cf.\ }}
\def\eg{{\it e.g.,\ }}
\def\ie{{\it i.e.,\ }}

\def\feh{{\rm[Fe/H]\ }}

\def\kms{{\rm\,km\,s^{-1}}}

\def\msun{{\rm\,M_\odot}}

\def\={\overline}

\def\lta{\mathrel{\spose{\lower 3pt\hbox{$\mathchar"218$}}
     \raise 2.0pt\hbox{$\mathchar"13C$}}}
\def\gta{\mathrel{\spose{\lower 3pt\hbox{$\mathchar"218$}}
     \raise 2.0pt\hbox{$\mathchar"13E$}}}
\def\Dt{\spose{\raise 1.5ex\hbox{\hskip3pt$\mathchar"201$}}}	
\def\dt{\spose{\raise 1.0ex\hbox{\hskip2pt$\mathchar"201$}}}	
\def\=={\equiv}

\def\dotsfill{\leaders\hbox to 1em{\hss.\hss}\hfill}

\def\sun{\odot}

\begin{document}

\title{The Stellar Populations of the Carina Dwarf Spheroidal Galaxy: \\
I. A New Color-Magnitude Diagram for the \\
Giant and Horizontal Branches}

\author{Tammy A. Smecker-Hane\altaffilmark{1,2},
Peter B. Stetson\altaffilmark{2}, James E. Hesser}
\affil{Dominion Astrophysical Observatory\\
Herzberg Institute for Astrophysics\\
National Research Council of Canada\\
5071 W. Saanich Rd., RR 5\\
Victoria, BC V8X 4M6 Canada}

\and

\author{Matthew D. Lehnert}
\affil{
Institute of Geophysics and Planetary Physics\\
Lawrence Livermore National Laboratory\\
7000 East Ave., L-413\\
Livermore, CA 94550 USA}

\altaffiltext{1}{NSF-NATO Post Doctoral Fellow}

\altaffiltext{2}{Visiting Astronomer, Cerro Tololo Inter-American
Observatory, which is operated by the Association of Universities for
Research in Astronomy, Inc. under contract with the National Science
Foundation.}

\begin{abstract}

We report on the first in a series of studies of the Carina dwarf
spheroidal galaxy, a nearby satellite of our Galaxy.  Our two major
results are: 1) precise $BI$ photometry ($\sigma_{B-I} \simlt 0.05$ for
$V \simlt 22$) for 11,489 stars in the Carina field, and 2) the
detection of two, morphologically distinct, horizontal branches in
Carina, which confirms that star formation occurred in two well-separated
episodes.  The old horizontal branch and RR Lyrae instability strip
belong to a $\simgt 10$~Gyr stellar population, while the populous
red-clump horizontal branch belongs, presumably, to a $\sim 6$~Gyr
stellar population.  We derive a distance modulus $(m-M)_0=20.09 \pm 0.06$
for Carina from the apparent magnitudes of the old horizontal branch and
the tip of the red giant branch (RGB), and discuss modifications to the
previously estimated distance, total magnitude, and stellar ages of
Carina.  Using the color of the RGB, we estimate the metallicities of
the younger and older populations to be $\feh \simeq -2.0$ and --2.2,
respectively.

\end{abstract}

\keywords{galaxies -- stellar populations}

\section{Introduction}

The dwarf spheroidal galaxies (dSphs) in the Local Group offer a unique
opportunity for studying stellar populations and testing models of
galactic evolution, because their size and proximity allow us to make
detailed measurements of their ages and metallicity distributions on a
star-by-star basis.  Color-magnitude diagrams and spectroscopic
abundance determinations for limited numbers of stars in the dSph
companions of the Galaxy show that each exhibits a different and, in
most cases, complex history of star formation. Some dSphs have
predominately old stellar populations with small ranges of age and
metallicity (e.g., Ursa Minor and Sculptor: Stetson 1984, Da Costa
1984), similar to Galactic globular clusters.  Others have
predominantly metal-poor, old populations, but still exhibit some
internal dispersion in age and/or metallicity (e.g., Draco: Zinn 1978,
Stetson 1984, Carney \& Seitzer 1984, Smith 1984, Lehnert \etal 1992).
However,  Carina and Fornax are examples of dSphs whose stellar
populations have a large range in both metallicity and age, and
evidence of a dominant intermediate-age ($\sim$6-9~Gyr) stellar
population (Carina: Mould \& Aaronson 1983 - Fornax: Aaronson \& Mould
1980, Buonanno \etal 1985, and references therein).

The Carina dwarf spheroidal galaxy\footnote{ Carina ($\alpha_{1950}=
{\rm 6^h 40.4^m}$, $\delta_{1950}= -50^\circ 55^{\prime}$;
$l=260.1^\circ$, $b=-22.2^\circ$) was discovered by Cannon \etal (1977)
from inspection of ESO/SRC Sky Survey plates.} is particularly
intriguing because it is suspected to have undergone two distinct
``bursts'' of star-formation. From the frequency of carbon stars, Mould
\& Aaronson (1983, hereafter MA83) found that a large fraction of Carina
stars were of intermediate age.  However, Saha \etal (1986, hereafter
SMS86) demonstrated the presence of an old stellar population by
detecting RR Lyrae stars in Carina.  Evidence for two distinct ages
and, hence, two bursts of star formation, also came from photometry of
the main-sequence turnoff (Mighell 1990).  Modelling two peaks detected
in the luminosity function of the main-squence turnoff region, Mighell
estimated that the first burst of star formation formed $\sim 17$\% of
Carina's stars $\sim 15$~Gyr ago, and the second burst formed the
majority of the stars ($\sim 83$\%) only $\sim7.5$~Gyr ago.  However,
the small size of the sample and comparatively large photometric errors
were enough to render this finding significant at only the $\sim 2
\sigma$ level. However, if such a long delay between two star-forming
events did indeed occur, then Carina presents some puzzles.  What
physical mechanism prevented the gas from cooling and forming stars in
the interim few Gyr, when we would expect the cooling timescale to be
of order 1~Gyr? How did the dark matter in Carina moderate its
evolution?

One possible explanation for the present appearance of the dSphs is
that supernovae drove large-scale galactic winds which halted star
formation, transforming dwarf irregulars into today's gas-poor dSphs
(\cf Dekel \& Silk 1986; Vader 1986; Silk, Wyse \& Shields 1987).  At
present, Carina is devoid of cold gas, with the upper limit on the mass
of HI being $10^3 \msun$ (Mould \etal 1990).  Massive dark-matter halos
inferred from velocity dispersion measurements (\cf Pryor 1992) may
have played a crucial role by providing stability and allowing the
stellar systems to remain intact even after copious gas loss.  Indeed,
the total mass-to-light ratio of Carina is inferred to be $M/L = 39 \pm
23$ from the 6.8 $\kms$ velocity dispersion (Mateo \etal 1993), which
suggests the presence of dark matter.  Dekel \& Silk (1986) show that
this formation hypothesis successfully predicts the observed
correlation between the mass-to-light ratio and the luminosity of dwarf
spheroidal and dwarf irregular galaxies. However, galactic winds may
{\it not\/} be required to explain the metal abundances of the
dwarf galaxies (Gilmore \& Wyse 1991),  while the existence of long
delays between {\it bursts} of star formation might argue {\it against}
galactic wind theories because dSphs are in low-density environments
and are unlikely to have accreted fresh gas.

To test theories of dynamical and chemical evolution of dwarf galaxies,
one must determine their star-formation history and metallicity
distributions.  Color-magnitude diagrams alone cannot unravel a
complicated star formation history because isochrones are degenerate in
distance, reddening, age, and metallicity, but spectroscopic chemical
abundances in concert with color-magnitude diagrams can.  Therefore, we
have begun a detailed study of Carina, which involves photometry of the
giant branch and main-sequence turnoff, as well as spectroscopy of
individual giants to determine the metal abundance distribution.  In
this first paper, we report on a color-magnitude diagram which contains
$\sim 95$\% of the giants in Carina. Two morphologically distinct
horizontal branches are seen, which supports the conclusion of two
discrete star-formation episodes in Carina.  In addition, we derive a
new distance modulus and a rough estimate of the metallicities of the
stars, and discuss the resulting revisions in the distance, stellar
ages, total magnitude and mass-to-light ratio of Carina.

\section{\bf Observations}

Direct images of Carina in $B$- and $I$-bands were obtained with the
CTIO 1.5-m telescope and TEK2048 CCD on 18--21 December 1992.  The
usable area of the CCD was $1830 \times 2040$ pixels due to vignetting,
and the pixel scale was 0.434$^{\prime\prime}$/pixel.  Photometric
conditions prevailed on the first two nights, and mild cirrus set in on
the third night.  Observations on the fourth night were interrupted by
intermittent cloud, although the data provided good relative photometry
within frames.  The mean seeing during the run was 1.5 arcseconds.  We
obtained 30 $B$-band and 42 $I$-band direct images in a pattern
covering the central 30\Min2$\times$24\Min6 of Carina, and in
additional fields along the major and minor axes. In total, the area
covered is roughly 47\% of the area enclosed within the tidal radius of
Carina. (For reference, Demers \etal [1983] derived $r_t =$ 33\Min8 and
$r_c =$ 10\Min7 by fitting King models to the surface density of
Carina.) Figure 1 shows a montage of the total area surveyed. Note that
there is significant contamination by foreground Galactic stars at this
latitude.

A typical sequence of observations at a given position consisted of
approximately 2 $\times$ 900s $B$ and 3 $\times$ 500s $I$ images, with
individual frames dithered by a few arcseconds in right ascension or
declination to move stars to different pixels and thereby average out
small-scale flat-fielding errors and contamination by detector
blemishes. Our mosaic pattern resulted in various total exposure times
for individual stars.  In the central body of Carina, the average
exposure time per star is approximately 2 hrs in each filter.  In the major
and minor-axis fields, the average exposure times were 45 and 30 min in the
$B-$ and $I$-bands, respectively.

Numerous standard-star fields from Landolt (1992) and Graham (1982) E
regions were observed throughout each night for photometric
calibration.  Shutter tests were made during the afternoons to map the
non-uniform exposure time across the TEK2048 CCD, which is a critical
correction required for high-precision photometry from short
integrations such as standard star exposures (Stetson 1989).  Other
standard calibration data included bias frames, and flat field
exposures taken with scattered sunlight reflected from the inside of
the dome, and twilight sky flats to correct for non-uniformities with
low spatial frequencies.

\section{\bf Data Reductions}

The standard CCD data reductions were preformed with personally written
software, and photometric data reductions were performed with ALLFRAME
(Stetson 1994). A mild non-linearity in the CCD was identified (see
Walker 1993) and satisfactory corrections applied to the calibrations.

\section{Results}

Figure 2 shows the color-magnitude diagram (CMD) of the 11,489 stars
observed in the Carina fields, and photometric standard errors as a
function of magnitude, and Figure 3 shows a schematic in which we have
identified the salient features of the CMD.  We have plotted an
approximate $V$ magnitude in order to display the CMD on a familiar
scale. To convert to $V$, we have calculated $(B-V)/(B-I)=0.48$ to be
the slope of the color-color relation applicable over most of our CMD
using photometry of the globular clusters NGC7006 and M92 courtesy of
L.  Davis (private communication to PBS). Evident in the CMD is a sheet
of foreground Galactic stars, which are primarily disk dwarfs.  A sharp
main-sequence turnoff in the thin/thick disk is prominent at
$(B-I)\approx 1.2$ for the full range of V magnitude.  Hereafter, we
concentrate on the Carina stars.

The most striking features of the Carina CMD are the {\it two,
morphologically distinct, horizontal branches} (HBs). The old HB with
blue and red extensions and an RR Lyrae strip is typical of old,
metal-poor globular clusters, \eg M15.  In fact, the old HB may indeed
have an extended blue HB, as does M15.  However, Carina also possesses
a red HB clump that is redder and more luminous than the old HB, and
has 2.6 times the number of stars, which suggests that the red HB clump
population is both more metal-rich and younger than the old HB
population (\eg Iben \& Rood 1970, Dorman 1992).

Another noteable feature of the CMD of Carina is the clearly defined
upper red giant branch (RGB) which is suprisingly thin, \ie comparable
to those of globular clusters.  Interestingly, the color distribution
of stars on the lower RGB shows tantalizing evidence for two stellar
populations.  To illustrate, we have calculated a cubic fit to the
locus of the RGB and plotted the distribution of stars away from the
mean locus, \ie $\delta_{B-I} \equiv (B-I) - (B-I)_{RGB}$, for stars
with $20.8 \leq V \leq 21.9$ (this range of $V$ avoids contamination by
HB and turnoff stars). Figure 4 shows the resulting histogram and the
best fitting model assuming a superposition of two Gaussian functions
and a sloping background. The dispersion of the Gaussian functions are
$\sigma = 0.062$ and 0.038 mags, which should be compared to the median
photometric standard errors for these stars which is $\sigma_{(B-I)} =
0.028$ mags.  The data appear to suggest the presence of two stellar
populations.  The first has an intrinsic dispersion in color that is a
factor of two larger than our photometric errors, which may signify an
intrinsic dispersion in metallicity as color in this region of the RGB
is highly sensitive to metallicity and only marginally sensitive to
age.  The second RGB appears to contain $\sim 10\%$ of the stars in
this region of the CMD, and it is approximately 0.16 mags bluer than
the first RGB. Are we resolving the older, more metal-poor stellar
population?  It is possible. However, we regard these tentative
conclusions as merely suggestive of further areas to examine when we
combine the present data with new, deeper photometry which we are
currently reducing.

The CMD also conains an additional sequence which is probably the
asymptotic giant branch (AGB) of the younger population (plausible because of
the appearence of a gap between the red HB clump and foot of the
sequence) and/or the RGB of the old population.  The clump of stars
appearing at the bottom of our CMD with $V\sim 22.5$ and $(B-I)\simeq
0.75$ is likely to be the tip of the main-sequence turnoff of the
younger population.

The dichotomy in the two HBs supports the conclusion reached by Mighell
(1990) --- from the double peaked color distribution in the
main-sequence turnoff region --- that star formation in Carina occurred
in two distinct bursts.  However, the appearance of two distinct HBs is
not conclusive evidence for two distinct stellar ages, because
metallicity and mass loss on the giant branch also play roles in HB
morphology.  We therefore hesitate to derive more quantitative
estimates of the ages of the two populations from the HBs.  The
evolution of the main-sequence turnoff is highly sensitive to age, and
better understood theoretically than evolution to the HB. Therefore, we
postpone detailed analysis of the CMD until we analyze the deep
photometry of the main-sequence turnoff region which we have obtained
with the CTIO 4m telescope (November 1993).  We will then compare these
CMDs with new theoretical isochrones spanning  the main-sequence
turnoff to the asymptotic giant branch, to constrain the ages and
metallicities of the stellar populations of Carina.

\subsection{Spatial Distribution of Carina Stars}

To illustrate the spatial distribution of Carina stars, we show in
Figure 5 the boundaries in the CMD which we have chosen to designate
probable Carina stars.  We used DAOPHOT to subtract the designated
foreground stars from our montage image (Figure 1) to show the Carina
dSph unveiled from most of the foreground contamination in Figure 6.

We can use the two distinct types of HBs stars as tracers of the young
and old stellar populations to test whether their spatial distributions
differ. From this we may be able to extract information about the
spatial extent of star formation in the two bursts.  For example, gas
dissipation could result in a second generation of stars that is more
centrally concentrated than the first. However, if the spatial
distributions of the two HB types are similar, then it is probable that
star formation in both bursts occurred over roughly the same areas,
because the relaxation timescale, $T_E$, in Carina is longer than a
Hubble time. To illustrate, the two-body relaxation timescale is
\[
 T_E = \frac{1}{16} \sqrt{\frac{3}{\pi}} \frac{v^3}
	{G^2 \rho m_2 \ln \Lambda} ,
\]
where
\[
\Lambda = \frac{Dv^2}{G (m_1+m_2)},
\]
(Chandrasekhar 1942)
for a body of mass $m_1$ in a density $\rho$ of particles of mass $m_2$
with velocity dispersion $v$ in a sytem of dimension $D$.  On the one
hand, the relaxation timescale due to stellar interactions in Carina is
approximately $4 \times 10^{12}$ yr assuming $m_1 = m_2 = 1 \msun$, $D
= $33\Min8 is the tidal radius (Demers \etal 1983), the central
velocity dispersion is 6.8 $\kms$ (Mateo \etal 1993), and the stellar
mass density is calculated using the central luminosity density of
$0.011 \msun pc^{-3}$  (Mateo \etal 1993) and a solar mass-to-light
ratio. On the other hand, one might appeal to massive dark matter
particles to stir the stars efficiently since the derived mass-to-light
ratio of Carina is 39 (Mateo \etal 1993).  To estimate the relaxation timescale
of the stars due to interactions with dark matter particles, we
estimate $\rho$ from the total mass of Carina which $1.1 \times 10^7
\msun$ (Mateo \etal (1993), and assume it is distributed uniformly over
a sphere of radius $D$.  In order for the relation timescale to be
shorter than $10^{10}$ yr, the dark matter must be composed of black
holes with mass $m \simgt 10^8 \msun$.  But if the dark matter in
Carina and the solar neighborhood are composed of similar objects, then
this is not a viable hypothesis, because the observed velocity
dispersion in the solar neighborhood places a limit of $m \simlt 10^6
\msun$ on dark matter particles (Lacey \& Ostriker 1985). Therefore,
lacking a mechanism to mix the two stellar populations in a few Gyr, if
the spatial distribution of the two stellar pupulations of Carina are
similar, then they must have formed over a similar spatial extent.

Thus we identified the old HB stars and the red clump HB stars from
their positions in the CMD. We examined the distribution of the stars
inside a limiting radius for which our sampling is complete (10
arcmin), which included 513 red clump HB stars and 185 old HB stars.
The surface densities as a function of radius for the two HB types
(normalizing the densities by the total number of stars of each type)
are shown in Figure 7. No significant difference in the surface
densities is detected.  In addition, we show in Figure 8 the fraction
of old HB stars as a function of radius, in which each radial bin was
chosen to contain 69 HB ($=$ old HB $+$ red HB clump) stars.  The mean
fraction of old HB stars is 0.27 and the most deviant point differs
only by $2 \sigma$ ($\sigma(\hbox{\rm one bin}) = 0.07$).  Again, no
systematic radial dependence is seen.  Therefore, we conclude that
inside 10 arcmin the radial distributions of the old and young stellar
populations are indistinguishable.  We infer that star formation in the
two bursts occurred with roughly the same spatial distribution inside
the core radius.

\section{Conclusions}

Here we use the absolute magnitude of the old HB and the absolute magnitude
of the tip of the red giant branch in our CMD to derive a distance
modulus to Carina, and the color of the red giant branch to estimate of
the metallicities of Carina stars.  We also discuss the resulting
modifications to the previously estimated distance, stellar ages,
total magnitude and mass-to-light ratio of Carina.

\subsection{Deriving a New Distance Modulus}

Past studies of Carina have usually assumed a distance modulus of
$(m-M)_0 =19.7$, which was derived by MA83. However, their derivation
was based on comparing the apparent magnitude of the {\it red clump} HB
to the absolute magnitude of {\it old} HBs/RR Lyrae stars in globular
clusters.  The  {\it old} HB of Carina was {\it not} detected by MA83,
because their small CCD size gave very limited spatial coverage of the
galaxy.  However, SMS86 obtained B-band photographic plates of Carina,
identified RR Lyraes, and derived $(m-M)_0=20.14$. Part of the reason
for the preference of the MA83 estimate in past studies may be that the
SMS86  data consisted of single-band photographic photometry, and they
misidentified as RR Lyraes a few bright variables which are anomalous
cepheids.  Our CMD resolves the discrepancy between these two distance
estimates, because it shows that the red HB clump is approximately 0.25
magnitudes brighter than the old HB.  Therefore, the correct distance
modulus of Carina is nearer to that derived by SMS86. Below, we derive
$(m-M)_0 =20.09 \pm 0.06$ with good agreement from two methods which
use the apparent magnitude of the old HB and the tip of the RGB.  In
the following, we will assume a reddening $E(B-V)=0.025$ and extinction
$A_V=0.08$ for Carina, as assumed by MA83 based on Burstein \& Heiles
HI data (1982, and 1983 private communication to MA).  The assumed
extinction laws are $A_I=1.98 E(B-V)$, $A_V=3.2 E(B-V)$, and $E(B-I) =
2.23 E(B-V)$ (Cardelli, Clayton \& Mathis 1989).

\subsubsection{Method 1: The Old HB/RR Lyrae Stars}

\label{sec-rrly}

The $M_V$ of the HB and RR Lyrae stars in old stellar populations has
been used extensively as a distance indicator. However, $M_V$ depends
on metallicity, and the exact dependence is controversial.  Theoretical
horizontal branch models of Lee, Demarque and Zinn (1990) give the
relation
\begin{equation}
M_V^{RR} = 0.17 \feh + 0.82.
\label{eqn-rrly}
 \end{equation}
If we assume that the metallicity of the old HB stars in Carina is
$\feh \simeq -2.2 \pm 0.4$, then $M_{V RR} = 0.45 \pm 0.07$. The
apparent magnitude of the old HB/RR Lyrae stars in Carina is $V=20.65
\pm 0.05$, and hence the derived distance modulus is $(m-M)_0=20.12 \pm
0.08$.  Additional spectroscopy of individual giants in Carina can
provide a metallicity distribution which will give us a firmer
constraint on this method for deriving the distance modulus.

\subsubsection{Method 2: The Tip of the RGB}

The absolute $I$ magnitude of tip of the red giant branch ($M_I^{TRGB}$) is
also a very good distance indicator (\cf Da Costa and Armandroff 1990,
Lee, Freedman \& Madore 1993, and references therein), because $M_I^{
TRGB} = -4.0$ with little variation for stellar populations with $\feh
< -0.7$, and ages $\simgt 7$~Gyr. By definition,
\[
M^{TRGB}_I =  M_{bol}^{TRGB} - BC_I,
\]
where $BC_I$ is the bolometric correction
in the $I$-band.  Da Costa and Armandroff (1990) give the parameteric fits:
\begin{eqnarray*}
     BC_I = 0.881 -0.243(V-I)_0^{TRGB}, \\
     M_{bol}^{TRGB} = -0.19 \feh - 3.81,
\end{eqnarray*}
derived from observed data on globular clusters (whose distances were
derived assuming Equation 1 above).

To apply this method to our data, we must estimate the slope of the
color-color relation applicable to the upper RGB.  We use the
semi-empirical calibration of the Revised Yale Isochrones (RYI, Green
\etal 1983) which was based on direct observations of globular cluster
stars. Considering isochrones for stars with ages $> 6$ Gyr and
metallicities $-1.7 \leq \feh \leq -2.3$, we adopt $(B-V)/(B-I) = 0.52
\pm 0.01$ for the upper RGB.  As a check, the slopes of the color-color
relation for the majority of our CMD from the RYI calibration and of
L. Davis' photometry of M92 and NGC7006 agree.

We find that it is straightforward to extract $(B-I)^{TRGB} = 2.85 \pm
0.05$ and $I^{TRGB} = 16.15 \pm 0.05$ by using either an eye-estimate
or by convolving the $I-$band luminisity function with a kernel used
for edge detection in image processing (\cf Lee \etal 1992). To
illustrate the strength of the edge of the TRGB, we plot in Figure 9
the $I$-band luminosity function for stars with $2.4 \leq (B-I) \leq
3.0$, which was chosen to limit contamination by foreground stars. If
we assume a metallicity of $\feh = -2.0 \pm 0.4$ (see below), then
$(V-I)^{TRGB}_0 = 1.45$ and $M^{TRGB}_I = -4.00$. Thus, the distance
modulus derived from the TRGB is $(m-M)_0 = 20.05 \pm 0.09$, which is
in good agreement with that derived above from the apparent magnitude
of old HB/RR Lyrae stars.

Therefore, we advocate a new distance modulus of $(m-M)_0 = 20.09 \pm
0.06$ for Carina. The revised heliocentric distance of Carina is 105
kpc, and the revised Galactocentric distance is 107 kpc, assuming
$R_\sun=8.5$ kpc.

Note that Mighell \& Butcher (1992) use their observed $V$-band
luminosity function to constrain the distance modulus (and
simultaneously the age and metallicity) of Carina from comparison with
the Revised Yale isochrones (Green \etal 1987).  Considering a range of
parameter space (in particular, distance moduli in bins of 0.2 mag),
they find models are ``consistent'' (defined by a reduced chi-square
signifying $>95$\% confidence) only with a distance modulus of 19.8.
Hence they estimate the distance modulus to be $19.8 \pm 0.1$.
However, we consider this conclusion uncertain for two main reasons:
1) the degeneracy of distance, age and metallicity in a CMD, and 2)
selective editing of their data.  In particular, Mighell \& Butcher
chose to delete an apparently discrepant data point at V=21.5 citing
possible contamination by foreground stars. However, if this point
where to be included in the analysis it would  significantly alter the
``knee'' in the luminosity function and only a single model would have
been found barely consistent with the data.  In fact, the inclusion of
that data point would favor either a larger distance modulus, lower
metallicity and/or younger age. Therefore, we do not think their result
of $19.8 \pm 0.1$ seriously conflicts with our determination of $20.09
\pm 0.06$ as the distance modulus.

\subsection{Estimating the Stellar Metallicity}

Da Costa and Armandroff (1990) show that the $(V-I)$ color
of the red giant branch at $M_I = -3$ is a sensitive indicator of
the metallicity of a stellar population. They derive the relation
\begin{eqnarray*}
     \feh = -15.16 + 17.0 (V-I)_{-3} - 4.9 (V-I)_{-3}^2 .
\end{eqnarray*}
We find $(B-I)_{-3} = 2.40$ for Carina, which gives $\feh = -2.18$ as
an estimate of the stellar metallicty  {\it if} the stars are as old as
globular clusters.

The giant branch (GB) of Carina is very thin, and the GBs of the old
and young populations appear to overlap (see Figure 4).  Thus, an
estimate of the metallicity of the older population is $\feh \simeq
-2.2$. However, in the following section, we revise the age of the
younger population to $\sim 6$~Gyr and must derive a correction for the
age difference.  Consulting Revised Yale Isochrones, we find that using
a 12~Gyr calibration for a 6~Gyr population results in a metallicity
estimate which is 0.22~dex too metal-poor. Hence our estimate of the
metallicity of young stellar population of Carina is $\feh = -2.0$.

Corroborating evidence for these metallicities comes from Da Costa \&
Hatzidimitriou (see Da Costa [1993]), who have obtained spectroscopic
metal abundances for 15 giants in Carina. They find a mean metallicity
of $\feh =-1.9$ with a very narrow spread, and one notable outlyer with
$\feh =-2.3$.

\subsection{Revised Stellar Ages}

Mighell (1990) identified two distinct main-sequence turnoffs in his
photometric data from which he estimated stellar ages.  He found the
turnoff for the younger population occurs at $V\approx 23.0$, and that
of the older population at $V\simgt 23.5$.
{}From VandenBerg \& Bell (1985) isochrones, Mighell derived
\[
\log t ({\rm Gyr}) = -0.13 \feh + 0.37 M_V^{MSTO} - 0.51.
 \]
If we assume our new distance modulus (0.3 mags greater than that
assumed by Mighell based on MA83) and $\feh = -2.0$ for the younger
stellar population, then the derived age is 6.2~Gyr, which is 17\%
younger than Mighell's original estimate. Assuming $\feh \simlt -2.2$
for the older population of Carina gives a lower limit of $\simgt
10$~Gyr for its age.

\subsection{Revised $M_V$ and $M/L$ Ratio}

Mateo \etal (1993) derived for Carina a total magnitude of $M_V = -8.9$
(based on the Demers \etal [1983] data) and a total mass-to-light ratio
of $M/L_V =39 \pm 23$ from their measured velocity dispersion and
derived structural parameters, assuming $(m-M)_0=19.7$ and $A_V=0.1$.
Adopting our new distance modulus and a smaller extinction instead
would lead to a total magnitude of $M_V = -9.3$ and a small decrease in
the derived mass-to-light ratio
\[
M/L_V = 39 \times 10^{(-0.2 \Delta (m-M)_0 - 0.4 \Delta A_V)} = 34.
\]

\section{Summary}

We have obtained precise photometry for most of the giant stars in
the Carina dSph galaxy and, for the first time, detect two
morphologically distinct horizontal branches, which most probably
represent two stellar populations with distinct ages and
metallicities.  We have derived a new distance modulus of $20.09 \pm
0.06$, and estimate the metallicities of the two populations (young and
old, respectively) to be $\feh = -2.0$ and --2.2. We propose revisions to
the stellar ages (6.2 and $>10$~Gyr, respectively), as
well as other fundamental parameters for Carina, such as its distance,
total magnitude and mass-to-light ratio. We await reduction of our
photometry of the main-sequence turnoff region of Carina, at which time
we will use stellar evolutionary isochrones and comparison globular
clusters of similar metallicity to refine the age estimates for
Carina's stellar populations, and to constrain betterthe duration of the
two bursts.  The final observations in this project will be
spectroscopy of $\sim 100$ individual giants identified in this survey
from which a metal abundance distribution can be derived.  Together,
these data will provide unprecedented constraints on the evolution of
the Carina dSph and insight into the evolution of dwarf galaxies.

\acknowledgments

We thank N. Suntzeff, A. Walker and R.  Schommer,  and our night
assistant, L. Gonzales, of CTIO for their help in obtaining these
observations.  As always, it was pleasure to use their excellent
facilities and support staff.  Lindsey Davis's private communication of
globular cluster photometry was greatly appreciated.  TSH acknowleges
the support of a NSF-NATO Postdoctoral Fellowship in 1993, and thanks
DAO for their hospitality and generous financial support.  The work of
MDL was performed at IGPP/LLNL under the auspices of the US Department
of Energy under contract W-7405-ENG-48.

\clearpage

\begin{figure}
\caption{Montage $B$-band image of the area surveyed around the Carina
dwarf spheroidal galaxy.  The central rectangular region is
30\Min2$\times$24\Min6, and the maximum dimensions of the region surveyed
are $41.5^\prime\times 33.3^\prime$. North is (approximately)
up and east is (approximately) left.}
\end{figure}

\begin{figure}
\caption{a) Color-magnitude diagram of all Carina and field stars, and b)
standard photometric errors returned from ALLFRAME.}
\end{figure}

\begin{figure}
\caption{Schematic illustration of the salient features of the CMD in
Figure 2.}
\end{figure}

\begin{figure}
\caption{The distribution of stars about the mean locus of the RGB
in the magnitude interval $20.8 \leq V 21.9$, and the best fitting
model assuming a superposition of two Gaussian functions and
a sloping background.}
\end{figure}

\begin{figure}
\caption{Color-magnitude diagram showing the adopted boundaries
for probable Carina giants.}
\end{figure}

\begin{figure}
\caption{Montage B-band image containing primarily Carina stars,
which was created by subtracting the foreground Galactic stars
identified in the making of Figure 4
from the original image (Figure 1).  North is (approximately)
up and east is (approximately) left.}
\end{figure}

\begin{figure}
\caption{Normalized surface density of the red clump HB stars
(solid line) and the old HB stars (dashed line) verses radius.}
\end{figure}

\begin{figure}
\caption{Fraction of old HB stars verses radius, in which
each bin was chosen to contain 69 HB stars.}
\end{figure}

\begin{figure}
\caption{The $I$-band luminosity function for stars
with $2.4 \leq (B-I) \leq 3.0$, which was chosen to limit contamination
by foreground stars.}
\end{figure}

\end{document}